\documentclass[preprint2]{aastex}
\usepackage{graphicx}
\usepackage[latin1]{inputenc}
\usepackage{txfonts}
\usepackage{lscape}
\usepackage{natbib}

\shorttitle{Decision Tree Classifiers for Star/Galaxy Separation}
\shortauthors{Vasconcellos et al.}

\begin{document}

\title{Decision Tree Classifiers for Star/Galaxy Separation}

\author{E. C. Vasconcellos$^{1}$, 
R. R. de Carvalho$^{2}$, R. R. Gal$^{3}$, F. L. LaBarbera$^{4}$,
H. V. Capelato$^{2}$, H. F. Campos Velho$^{5}$,  M. Trevisan$^{6}$ and R. S. R. Ruiz$^{1}$\\
$^{1}$CAP, National Institute of Space Research, Av. dos Astronautas 1758, S$\tilde{a}$o Jos\'e dos Campos 12227-010, Brazil\\
$^{2}$DAS, National Institute of Space Research, Av. dos Astronautas 1758, S$\tilde{a}$o Jos\'e dos Campos 12227-010, Brazil\\
$^{3}$Institute for Astronomy, University of Hawaii, 2680 Woodlawn Dr., Honolulu, HI 96822, United States\\
$^{4}$Osservatorio Astronomico di Capodimonte, Nacional Institute for Astrophysics, via Moiariello 16, Napoli 80131, Italy\\
$^{5}$LAC, National Institute of Space Research, Av. dos Astronautas 1758, S$\tilde{a}$o Jos\'e dos Campos 12227-010, Brazil\\
$^{6}$IAG, University of São Paulo, Rua do Matão 1226, S$\tilde{a}$o Paulo 05508-090, Brazil}

\begin{abstract}
  We study the star/galaxy classification efficiency of 13 different
  decision tree algorithms applied to photometric objects in the Sloan
  Digital Sky Survey Data Release Seven (SDSS DR7). Each algorithm is
  defined by a set of parameters which, when varied, produce different
  final classification trees.  We extensively explore the parameter
  space of each algorithm, using the set of $884,126$ SDSS objects
  with spectroscopic data as the training set.  The efficiency of
  star-galaxy separation is measured using the completeness function.
  We find that the Functional Tree algorithm (FT) yields the best
  results as measured by the mean completeness in two magnitude
  intervals: $14\le r\le21$ ($85.2\%$) and $r\ge19$ ($82.1\%$).  We
  compare the performance of the tree generated with the optimal FT
  configuration to the classifications provided by the SDSS parametric
  classifier, 2DPHOT and Ball et al.  (2006).  We find that our FT
  classifier is comparable or better in completeness over the full
  magnitude range $15\le r\le21$, with much lower contamination than
  all but the Ball et al. classifier. At the faintest magnitudes
  ($r>19$), our classifier is the only one able to maintain high
  completeness ($>$80\%) while still achieving low contamination
  ($\sim2.5\%$).  Finally, we apply our FT classifier to separate
  stars from galaxies in the full set of $69,545,326$ SDSS photometric
  objects in the magnitude range $14\le r\le21$.
\end{abstract}


\section{Introduction}
\label{intro}

Astronomical data acquisition has experienced a revolution both in quality and complexity during the last three decades. The main driver has been the deployment of, and enormous growth in, modern digital CCD detectors that replaced venerable photographic plates in the 1980s. Digital images provided by CCDs, coupled with rapid developments in computation and data storage, made it possible and even routine to produce terabytes of astrophysical data in a year. Moreover, several large scale surveys are being planned for the next ten years, generating a vast quantity of deep and wide photometric images. These surveys will provide data at rates and volumes much greater than any previous projects. Therefore it is necessary to not only develop new methods for processing and analyzing such huge data volumes, but also to ensure that the techniques applied to extract information from the data are optimal.

A basic step in the extraction of sensible astronomical data from photometric images is separating intrinsically pointlike sources (stars) from extended ones (galaxies). Distinguishing between these two classes becomes increasingly difficult as sources become fainter due to the lack of spatial resolution and signal-to-noise. Our goal in this work is to test a variety of decision tree classifiers, and ultimately perform reliable star/galaxy separation for objects from the Seventh  Data Release of the Sloan Digital Sky Survey (SDSS-DR7; Abazajian et al. 2009) based on photometric data. We use the SDSS because it also contains an enormous number of objects with spectroscopic data (which give the true object classes), and because of the quality, consistency and accuracy of its photometric data.

In the 1970s and 1980s, when digitized images became widespread in astronomy, many authors undertook projects to create automated methods to separate stars from galaxies. The first efforts relied on purely parametric methods, such as the pioneering works of Macgillivray et al. (1976), Heydon-Dumbleton et al. (1989) and Maddox et al. (1990). Macgillivray et al. (1976) used a plot of transmission vs. $\log{\left(\mbox{area}\right)}$\footnote{Decimal logarithm of occupied area of the object in the image. The area is measured as the number of squares with side equal to $8\mu $m}, fitting a discriminant function to separate stars and galaxies. Their star/galaxy separation had a completeness (i.e, the fraction of all galaxies classified as such) of $95\%$ and a contamination (fraction of non-galaxy objects classified as galaxies) of $5$-$10\%$. Heydon-Dumbleton et al. (1989) performed star/galaxy separation on 200 photographic plates digitized by COSMOS. Rather than use directly measured object attributes, they generated classification parameters based on the data, plotting these as a function of magnitude in bi-dimensional parametric diagrams. In these classification spaces they  then used an automated procedure to derive separation functions. They reached $98\pm2\%$ completeness with $8\pm2\%$ contamination. Maddox et al. (1990) used a set of 10 parameters measured by APM in 600 digitized photographic plates from the UK Schmidt Telescope. They reached $90\%$ completeness with $10\%$ contamination at magnitudes $B_j < 20.5$. All of these completeness and contamination must be treated with caution, as they are based on comparison to expected number counts and plate overlaps, rather than a spectroscopic ``truth'' sample.

As the volume of digital data expanded, along with the available computing power, many authors began to apply machine learning methods like decision trees (DT - see below, Section \ref{DT}) and neural networks to address star/galaxy separation. Unlike parametric methods, machine learning methods do not suffer from the subjective choice of discriminant functions and are more efficient at separating stars from galaxies at fainter magnitudes (Weir et al. 1995). These methods can incorporate a large number of photometric measurements, allowing the creation of a classifier more accurate than those based on parametric methods. Weir et al. (1995) applied two different DT algorithms, the GID*3 (Fayyad, 1994) and the O-Btree (Fayyad \& Irani, 1992) as star/galaxy separators for images from the Digitized Second Palomar Observatory Sky Survey (DPOSS), obtaining $90\%$ completeness and $10\%$ contamination. Odewahn et al. (1999) applied a neural network to DPOSS images and established a catalog spanning 1000 square degrees. Odewahn et al. (2004) used a DT and a neural network to separate objects in DPOSS and found that both methods have the same accuracy, but the DT consumes less time in the learning process. Suchkov et al. (2005) was the first to apply a DT to separate objects from the Sloan Sky Digital Survey (SDSS). The authors applied the oblique decision tree classifier ClassX, based on OC1, to the SDSS-DR2 (Abazajian et al. 2004). They classified objects into stars, red stars (type M or later), AGN and galaxies, giving a percentage table of correct classifications that allows one to estimate their completeness and contamination. Ball et al. (2006) applied an axis-parallel decision tree. These authors used $477,068$ objects from SDSS-DR3 (Abazajian et al. 2005) to build the decision tree~-~the largest training set ever used. They obtained a completeness of $93.8\%$ for galaxies and $95.4\%$ for stars.

In this paper we employ a DT machine learning algorithm to separate objects from SDSS-DR7  into stars and galaxies. We evaluate 13 different DT algorithms provided by the WEKA (Waikato Environment for Knowledge Analysis) data mining tool. We use a training data set containing only objects with measured spectra. The algorithm with the best performance on the training data was then used to separate objects in the much larger data set of objects having only photometric data. This is the first work published testing such a large variety of algorithms and using all of the data in the final SDSS data release (see also Ruiz et al. 2009).

Improving star/galaxy separation at the faintest depths of imaging surveys is not merely an academic exercise. By significantly improving the completeness in faint galaxy samples, and reducing the contamination by misclassified stars, many astrophysically important questions can be better addressed. Mapping the signature of  baryon acoustic oscillations requires large galaxy samples - and the more complete at higher redshift, the better.  The measurement of galaxy-galaxy correlation functions is of course improved, both by increasing the number of galaxies used, and by reducing the washing out of the signal due to the smooth distribution of erroneously classified stars. Weak lensing surveys, which need the largest and purest sample of background (lensed) galaxies and excellent photometric redshifts benefit on both fronts. Similarly, searches for galaxy clusters using galaxy overdensities increase their efficiency when there are fewer contaminant stars and more constituent galaxies. Searches for rare objects, both stellar and extended, also win with reduced contamination, as do any programs which target objects for follow-up spectroscopy based on the source type.

For future imaging surveys, optimized classifiers will require a new breed of training set. Because they cover large sky areas, programs like the Dark Energy Survey (DES), the Large Synoptic Survey Telescope (LSST) and Pan-STaRRS can utilize all available spectroscopy to create training samples, even to quite faint magnitudes. Because most spectroscopy has targeted galaxies, the inclusion of definite stars must be accomplished in another way. Hubble Space Telescope (HST) images have superb resolution and can be used to determine the morphological class (star or galaxy) of almost all objects observed by HST. Although covering only a tiny fraction of the sky, the depth of even single orbit HST images and the area overlap with these large surveys will provide star/galaxy (and perhaps even galaxy morphology) training sets that are more than sufficient to implement within an algorithm like the one we describe.

The structure of this paper is as follows. In \S ~\ref{Data} we describe the SDSS data used to evaluate the WEKA algorithms. In \S ~\ref{DT} we give a brief description of the DT method and discuss the technique used to choose the best WEKA decision tree building algorithm. In \S ~\ref{separation} we discuss the evaluation process and the results for each algorithm tested. In \S ~\ref{comparativestudy} we compare our best star/galaxy separation method to the  SDSS parametric method (York et al. 2000), the 2DPHOT parametric method (La Barbera et al. 2008)  and the axis-parallel DT used by Ball et al. (2006). We also examine whether the SDSS parametric classifier can be improved by modifying the dividing line between stars and galaxies in the classifier's parameter space. We summarize our results in \S \ref{summary}.

\section{The Data}
\label{Data}

We used simple Structured Query Language (SQL) queries to select data from the SDSS Legacy survey\footnote{These queries were written for the SDSS Sky Server database which is the DR7 Catalog Archive Server (CAS); see 
http://cas.sdss.org/astrodr7/en/. Photometric data was obtained through the  \texttt{photoObj} view of the database and spectroscopic data through the \texttt{specObj} view.}.  Objects were selected having $r$-magnitudes in the range 14$^{m}$- 21$^{m}$. We obtained two different data samples: the {\it spectroscopic}, or {\it training}, sample and the {\it application} sample. The spectroscopic sample (see \S ~\ref{Attributes}) contains only those objects with both photometric and spectroscopic measurements, while objects in the application sample have only photometric measurements. The spectroscopic sample was obtained through the following query: 

\noindent \hspace*{0.5cm} \texttt{SELECT} \\
   \hspace*{1cm} \texttt{p.objID, p.ra, p.dec, s.specObjID,} \\
   \hspace*{1cm} \texttt{p.psfMag\_r, p.modelMag\_r, p.petroMag\_r,} \\
   \hspace*{1cm} \texttt{p.fiberMag\_r, p.petroRad\_r, p.petroR50\_r,} \\
   \hspace*{1cm} \texttt{p.petroR90\_r, p.lnLStar\_r,p.lnLExp\_r,} \\
   \hspace*{1cm} \texttt{p.lnLDeV\_r, p.mE1\_r, p.mE2\_r, p.mRrCc\_r,} \\
   \hspace*{1cm} \texttt{p.type\_r,p.type, s.specClass} \\
\hspace*{0.5cm} \texttt{FROM PhotoObj AS p} \\
   \hspace*{1cm} \texttt{JOIN SpecObj AS s ON s.bestobjid = p.objid} \\
\hspace*{0.5cm} \texttt{WHERE} \\
   \hspace*{1cm} \texttt{p.modelMag\_r BETWEEN 14.0 AND 21.0} \\

\noindent This query returned slightly over one million objects assigned to six different classes according to their SDSS spectral class\footnote{For more information about spectral class please refer to \S~\ref{Attributes}.}. However, only objects of spectral class star and galaxy are used, leaving us with 884,378 objects for the spectroscopic sample. The majority of excluded objects are spectroscopically QSOs (9.1\% of the query results), many of which have one or more saturated pixels in the photometry.
We also removed 51 stars and 147 galaxies with non-physical values (e.g. -9999) for some of their photometric attributes. Finally we excluded 54 objects found to be repeated SDSS spectroscopic targets, leaving a final training sample of $884,126$ objects, consisting of $84,043$ stars and $800,083$ galaxies. These all have reliable SDSS star or galaxy spectral classifications and meaningful photometric attributes.

The application sample was built similarly to the spectroscopic sample with the following query:

\noindent \hspace*{0.5cm} \texttt{SELECT} \\
   \hspace*{1cm} \texttt{objID, ra, dec, psfMag\_r, modelMag\_r,} \\
   \hspace*{1cm} \texttt{petroMag\_r, fiberMag\_r, petroRad\_r,} \\
   \hspace*{1cm} \texttt{petroR50\_r, petroR90\_r, lnLStar\_r,} \\
   \hspace*{1cm} \texttt{lnLExp\_r, lnLDeV\_r, mE1\_r, mE2\_r,} \\
   \hspace*{1cm} \texttt{mRrCc\_r, type\_r, type} \\
\hspace*{0.5cm} \texttt{FROM PhotoObj} \\
\hspace*{0.5cm} \texttt{WHERE} \\
   \hspace*{1cm} \texttt{modelMag\_r BETWEEN 14.0 AND 21.0} \\
   
\noindent This retrieved photometric data for nearly 70 million objects from the Legacy survey. We use the Legacy survey rather than SEGUE because we are interested in classifying distant objects at the faint magnitude limit of the SDSS catalog.


\section{The Decision Tree Method}
\label{DT}

Machine learning methods are algorithms that allow a computer to distinguish between classes of objects in massive data sets by first ``learning'' from a fraction of the data set for which the classes are known and well defined~-~ the {\it training} set. Machine learning methods are essential to search for potentially useful information in large, high-dimensional data sets. 

\begin{figure}
\centering
\includegraphics[width=70mm,height=45mm]{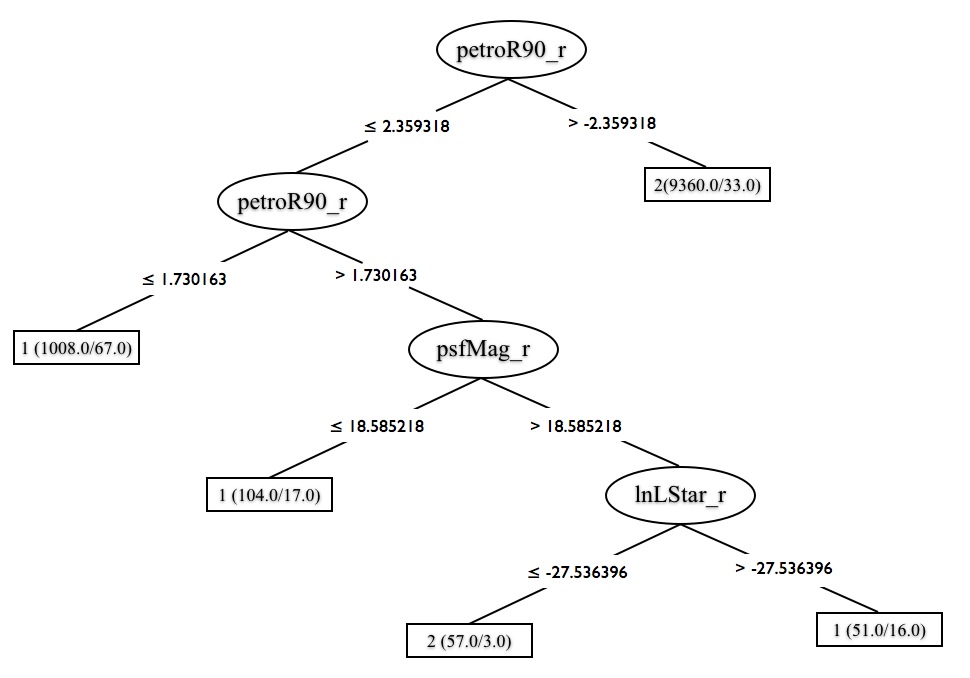}
\caption{A simple decision tree built with the J48 algorithm (Witten \& Frank, 2000). This tree was trained with 50,000 objects from the spectroscopic sample, as described in \S~\ref{Data}, and has a minimum number of objects per leaf equal to 50.}
\label{fig.simpletree}
\end{figure}

A DT is a well-defined machine learning method consisting of nodes which are simple tests on individual or combined data attributes. Each possible outcome of a test corresponds to an outgoing branch of the node, which leads to another node representing another test and so on. The process continues until a final node, called a leaf, is reached. Figure \ref{fig.simpletree} shows a graphical representation of a simple DT constructed with $50,000$  randomly chosen SDSS objects having spectroscopic data. At its topmost node (the root node) the tree may branch left or right depending on whether the value of the data attribute petroR90 is less than or greater than $2.359318$. Either of these branches may lead to a child node which may test the same attribute, a different one, or a combination of attributes. The path from the root node to a leaf corresponds to a single classification rule.

Building up a DT is a supervised learning process, i.e., the DT is built node by node based on a data set where the classes are already known. This dataset is formed from training examples, each consisting of a combination of attribute values that leads to a class. The process starts with all training examples in the root node of the tree. An attribute is chosen for testing in the node. Then, for each possible result of the test a branch is created and the dataset is split into subsets of training examples that have the attribute values specified by the branch. A child node is created for each branch and the process is repeated, splitting each subset into new subsets. Child nodes are created recursively until all training examples have the same class or all the training examples at the node have the same values for all the attributes. Each leaf node gives either a classification, a set of classifications, or a probability distribution over all possible classifications. The main difference between the different algorithms for constructing a DT is the methood(s) employed to select which attribute or combination of attributes to be tested in a node.


\subsection{WEKA and Tree Construction}
\label{weka}

WEKA\footnote{http://www.cs.waikato.ac.nz/ml/weka/} is a Java Software package for data mining tasks developed by the University of Waikato, New Zealand. It consists of a collection of machine learning algorithms that can either be  applied directly or called from an another Java code. 
WEKA contains tools for data pre-processing, classification, regression, clustering, association rules, and visualization.

In this work we use the WEKA DT tools, which include 13 different and independent algorithms for constructing decision trees\footnote{There are other DT algorithms in WEKA which are not capable of working with numerical attributes.}. 
In the following we give a brief description of each algorithm.
\begin{description}
\item[J48] is the WEKA implementation of the C4.5 algorithm (Quinlam, 1993). Given a data set, it generates a DT  by recursive partitioning of the data. The tree is grown using a depth-first strategy, i.e., the algorithm calculates the information gain for all possible tests that can split the data set and selects a test that gives the greatest value. This process is repeated for each new node until a leaf node has been reached.
\item[J48graft] generates a grafted DT from a J48 tree. The grafting technique (Webb, 1999) adds nodes to an existing DT with the purpose of reducing prediction errors. This algorithm identifies regions of the multidimensional space of attributes not occupied by training examples, or occupied only by misclassified training examples, and considers alternative branches for the leaf containing the region in question. In other words, a new test will be performed in the leaf, generating new branches that will lead to new classifications.
\item[BFTree] (Best-First decision Tree; Haijian Shi, 2007) has a construction process similar to C4.5. The main difference is that C4.5 uses a fixed order to build up a node (normally left to right), while BFTree uses the best-first order. This method first builds the nodes that will lead to the longest possible paths (a path is the way from the root node to a leaf).
\item[FT] (Functional Trees; Gama, 2004) combines a standard univariate DT, such as C4.5, with linear functions of the attributes by means of linear regressions. While a univariate DT uses simple value tests on single attributes in a node, FT can use linear combinations of different attributes in a node or in a leaf.
\item[LMT] (Logistic Model Trees, Landwher et al. 2006) builds trees with linear functions in leafs as does the FT algorithm. The main difference is that instead of using linear regression, LMT uses logistic regression.
\item[Simple Cart] is the WEKA implementation of the CART algorithm (Breiman et al. 1984). It is similar to C4.5 in the process of tree construction, but while C4.5 uses information gain to select the best test to be performed on a node, CART uses the Gini index.
\item[REPTree] is a fast decision tree learner that builds a decision/regression tree using information gain/variance as the criterion to select the attribute to be tested in a node.
\item[Random tree] models have been extensively developed 
in recent years. The WEKA Random Tree algorithm builds a tree considering K randomly chosen attributes at each node.
\item[Random Forest] (Breiman, 2001) generates an ensemble of trees, each built from random samples of the training set. The final classification is obtained by majority vote.
\item[NBTree] (Naive Bayesian Tree learner algorithm; Kohavi, 1996) generates a hybrid of  a Naive-Bayesian classifier and a DT classifier. The algorithm builds up a tree in which the nodes contain univariate tests, as in a regular DT, but the leaves contains Naive-Bayesian classifiers. In the final tree an instance is classified using a local Naive Bayes on the leaf in which it fell. NBTree frequently achieves higher accuracy than either a Naive Bayesian classifier or a DT learner.
\item[ADTree] (Alternating Decision Tree; Freund and Mason, 1999) is a boosted DT. An ADTree consists of prediction nodes and splitter nodes. The splitter nodes are defined by an algorithm test, as, for instance, in C4.5, whereas a prediction node is defined by a single value $x \in R^2$. In a standard tree like C4.5 a set of attributes will follow a path from the root to a leaf according to the attribute values of the set, with the leaf representing the classification of the set. In an ADTree the process is similar but there are no leaves. The classification is obtained by the sign of the sum of all prediction nodes existing in the path. Different from standard trees, a path in an ADTree begins at a prediction node and ends in a prediction node.
\item[LADTree] (Holmes et al. 2001) produces an ADTree capable of dealing with data sets containing more than 2 classes. The original formulation of the ADTree restricted it to binary classification problems; the LADTree algorithm extends the ADTree algorithm to the multi-class case by splitting the problem into several two-class problems.
\item[Decision Stump] is a simple binary DT classifier consisting of a single node (based on one attribute) and two leaves. All attributes used by the other trees are tested and the one giving the best classifications (PetroR50 in our case) is chosen to use in the single node.
\end{description}


\subsection{Accuracy and Performance: the Cross-Validation Method}
\label{cross}

The accuracy of any method for star/galaxy separation depends on the apparent magnitude of the objects and is often measured using the Completeness function $CP(m)$ (fraction of all galaxies classified as such)  and the Contamination function $CT(m)$ (fraction of all stars classified as galaxies) in a magnitude interval $\delta m$. These are defined as:
\begin{eqnarray}
CP(m) &=& 100\ast \frac{N_{gal-gal}(m)\delta m}{N_{galaxy}^{tot}(m)\delta m} \\
\mathrm {and}& & \nonumber\\
CT(m) &=&  100\ast \frac{N_{star-gal}(m)\delta m}{N_{star}^{tot}(m)\delta m} 
\end{eqnarray}
where $N_{gal-gal}(m)\delta m$ is the number of galaxy images correctly identified as galaxies within the magnitude interval $(m-\delta m /2,m+\delta m /2)$;  $N_{star-gal}(m)\delta m$ is the number of star images falsely identified as galaxies; $N_{galaxy}^{tot}(m)\delta m$ is the total number of galaxies and $N_{star}^{tot}(m)\delta m$  the total number of stars. 

It is also useful to define the mean values of these functions within a given magnitude interval. Thus for the mean completeness we have: 
$\langle\mathrm{Compl}\rangle_{\mathrm{\Delta m}} = (1/ \Delta m)\sum{CP(m_{i}) \delta m_{i}}$, with $\Delta m = \sum{\delta m_{i}}$. A similar definition holds for the mean contamination. Note that $\langle\mathrm{Compl}\rangle_{\mathrm{\Delta m}} \cdot \Delta m$ gives the area under the completeness function in the interval $\Delta m$. Unless otherwise stated, we calculate the completeness and contamination functions using a constant bin width $\delta m = 0.5^{m}$.

Our first goal is to find the best performing DT algorithm among those described in Section \ref{weka}, in terms of accuracy, especially at faint magnitudes. However, for large data sets, the processing time is also a concern in this evaluation. 

There are various approaches to determining the performance of a DT. The most common approach is to split the training set into two subsets, usually in a 4:1 ratio, and construct a tree with the larger subset and apply it to the smaller. A more sophisticated method is called Cross-Validation (CV; Witten \& Franck, 2000). The CV method, which is used here, consists of splitting the training set into 20 subsamples, each with the same distribution of classes as the full training set. While the number of subsamples, 20, is arbitrary, each subsample must provide a large training
set for the CV method. For each subsample a DT is built and applied to the other 19 subsamples. The resulting completeness and contamination functions are then collected and the median and dispersion over all subsets is found. This gives the cross-validation estimate of the robustness in terms of a completeness function.

\section{Star/Galaxy Separation for the SDSS}
\label{separation}

The spectroscopic information provided by SDSS provides the true classification~-~star or galaxy~-~of an object. Despite the size of the SDSS spectroscopic sample ($\sim1$ million objects), it represents only a tiny fraction of all objects in the SDSS DR7 photometry ($230$ million). How can we classify the SDSS objects for which there is no spectroscopic data? The SDSS pipeline already produces a classification using a parametric method based on the difference between the magnitudes \texttt{psfMag} and \texttt{modelMag} (see \S ~\ref{Attributes}). However it is known (see Figure 6 below) that this classification is not very accurate at magnitudes fainter than $19.0$. 

We will take advantage of the large spectroscopic sample from SDSS, for which we know the correct classes for all objects, to train a DT to classify SDSS objects based only on their photometric attributes. We expect that by using such a vast training set, the resulting DT will be capable of maintaining good accuracy even at faint magnitude.

\subsection{Attributes}
\label{Attributes}

We selected 13 SDSS photometric attributes and a single spectroscopic attribute (\texttt{specClass}), as shown in Table \ref{tab.attributes}.

\begin{table}
\caption{SDSS-DR7 attributes used for star/galaxy separation.} \label{tab.attributes}
\begin{center}
{\scriptsize\begin{tabular}{ll}\hline \hline
\multicolumn{1}{c}{Attribute} & \multicolumn{1}{c}{CAS Variable}\\ \hline
PSF Magnitude & \texttt{psfMag} \\
Fiber Magnitude & \texttt{fiberMag} \\
Petrosian Magnitude & \texttt{petroMag} \\
Model Magnitude & \texttt{modelMag} \\
Petrosian Radius & \texttt{petroRad} \\
Radius carrying 50\% of petrosian flux & \texttt{petroR50} \\
Radius carrying 90\% of petrosian flux & \texttt{petroR90} \\
Likelihood PSF & \texttt{lnLStar} \\
Likelihood Exponential & \texttt{lnLExp} \\
Likelihood deVaucouleurs & \texttt{lnLDeV} \\
Adaptive Moments & \texttt{mRrCc}, \texttt{mE1} e \texttt{mE2} \\
Spectroscopic classification & \texttt{specClass} \\
\hline \hline
\end{tabular}}
\end{center}
\end{table}

This set of  photometric attributes is the same for both the spectroscopic (training) and the application samples.  While one could ask what set of attributes produces the most accurate star/galaxy separation, the enormous variety of attributes measured by SDSS for each photometric object places examination of that question beyond the scope of this work. We instead focus on those attributes that are known or expected to strongly correlate with the object classification. These attributes are:\\
\begin{itemize}
\item{The {\it PSF magnitude} (\texttt{psfMag}), described in detail in Stoughton et al. (2002), obtained by fitting a Point Spread Function (PSF) Gaussian model to the brightness distribution of the object. 
We expect the PSF magnitude to be a good flux measure for stars, but it tends to overestimate  the flux of extended objects due to their irregular shapes. }
\item{The {\it fiber magnitude} (\texttt{fiberMag}) is the flux contained within the 3'' diameter aperture of a spectroscopic fiber.}
\item{The {\it petrosian magnitude} (\texttt{petroMag}) is a flux measure proposed by Petrosian (1976). He defined a function $\eta(r)$ representing the ratio between the mean surface brightness within a specific radius and the surface brightness at this radius. For a given value of $\eta$ one can define a {\it petrosian radius} (\texttt{petroRad}); the flux measured within this radius is the petrosian magnitude. Note that the SDSS pipeline adopts a modified form of the
Petrosian (1976) system, as detailed in Yasuda et al. (2001).}
\item{The SDSS pipeline fits two different galaxy models to the two-dimensional image of an object, in each band: a de Vaucouleurs profile and an exponential profile. The {\it model magnitude} (\texttt{modelMag}) is taken from the better fitting of these two models\footnote{For more details see http://www.sdss.org/dr7/algorithms photometry.html}.
}.  
\item{The attributes \texttt{petroR50} and \texttt{petroR90} are the radii containing $50\%$ and $90\%$ of the Petrosian flux for each band. These two attributes are not corrected for seeing and this may cause the surface brightness of objects of size comparable to the PSF to be underestimated. Nevertheless, the amplitude of these effects are not yet well characterized, and machine learning algorithms may still find relationships distinguishing stars from galaxies.}
\item{The {\it model likelihoods} \texttt{lnLStar, lnLExp} and \texttt{lnLDeV} are the probabilities that an object would have at least the measured value of chi-squared if it were well represented by one of the SDSS surface brightness models: PSF, de Vaucouleurs or exponential, respectively. }
\item{ The {\it Adaptive moments} \texttt{mRrCc,mE1} and \texttt{mE2} are second moments of the intensity, measured using a radial weight function adapted to the shape and size of an object. A more detailed description can be found in Bernstein \& Jarvis (2002). Adaptive moments can be a good measure of the ellipticity. }
\item{The spectroscopic attribute \texttt{specClass} stores the object spectral classification, which is one of \texttt{unknown,star,galaxy,qso,hiz\_qso,sky,star\_late} or \texttt{gal\_em}. Only objects classified as stars or galaxies are kept in the training set. All other spectroscopic classes constitute a small fraction of the objects, and are not uniquely tied to a specific morphological class.}
\end{itemize}

\begin{figure*}
\centering
\includegraphics[width=120mm,height=190mm]{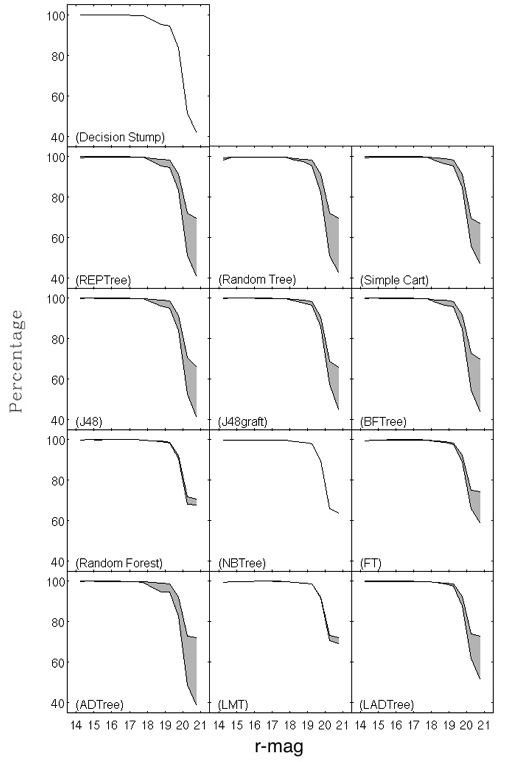}
\caption{Results of the parameter space exploration for each of the 13 WEKA DT algorithms. The hatched areas show the loci of the completeness functions for galaxies obtained from the CV procedure.}
\label{fig.parameter_space}
\end{figure*}

\subsection{Objective Selection of the Optimal Tree Algorithm}
\label{tests}

As discussed in Section \ref{weka}, the WEKA data mining package
provides 13 different algorithms to generate a DT from a training set
containing only numerical attributes. Each algorithm employs different
computational procedures using different sets of internal parameters
\footnote{We do not provide a full description of all the
  parameters involved in each algorithm; for more
  detailed descriptions please refer to
  http://www.cs.waikato.ac.nz/ml/weka/} resulting in construction of
distinct final trees. For each algorithm, we test various sets of
internal parameters (always with the same input object
attributes). We use the CV procedure (Section \ref{cross}) to compute
the completeness function for each algorithm and all sets of its
internal parameters, to find the best set of parameters maximizing the
completeness. Then, we compare the perfomance among the various
algorithms using the optimal internal parameters for each algorithm.
In this section we discuss these tests and compare their results to
find the optimal algorithm (and its best internal parameters) that
will ultimately be used to provide star-galaxy separation for the
entire DR7.

We first exhaustively explored the internal parameter space of each
algorithm to determine which parameters significantly change the
resulting completeness functions, and discarded the irrelevant
ones\footnote{Note that the Decision Stump and NBTree have no free
  parameters} . We tested the sensitivity of the completeness function
to the variation of each parameter taken individually (``single
tests'') or in combination with others (``combined tests''). For each
set of completeness functions generated by the variation of a given
parameter $\nu$, we computed the dispersion $\sigma_{m_{i}}$ at the
middlepoints $m_{i} = 14.25^{m} + i \ast 0.5^{m}$ and then averaged
over all the intervals to obtain $\sigma_{\nu} = (1/N_{intervals})
\sum \sigma_{m_{i}})$.  A parameter was considered irrelevant whenever
$\sigma_{\nu} \le 5\%$. This procedure typically allowed us to discard
one parameter per algorithm.

In the second step we searched for the optimal value of each remaining
parameter. To do this we first computed the range for each parameter
wherein $\sigma_{\nu} \le 5\%$, as before. Then, within these limits we
tested each algorithm to find its optimal parameter set. The results
of these tests are shown in Figure 2, which
displays the range of completeness functions computed using the CV
procedure when varying the internal parameters of each algorithm. At
bright magnitudes, $r \lesssim 19$, the algorithms behave very
similarly, and their efficiency is stable under variation of their
internal parameters.

We then analyze the relative performance of the 13 algorithms by
comparing their completeness and contamination functions as well as
their processing times when run with their optimal sets of internal
parameters.  We define the quantities
$\langle\mbox{Compl}\rangle_{\mbox{\scriptsize{bright}}}$,
$\langle\mbox{Compl}\rangle_{\mbox{\scriptsize{faint}}}$ and
$\mbox{Compl}_{20.75}$ which are the mean completenesses (see Section
\ref{cross}) in the magnitude intervals $14\le r < 19$ , $19\le r \le
21$ and $20.5\le r \le 21$, respectively. The results of this
comparison are given in Table \ref{tab.parameter_space}. Column
1 gives the algorithm name as in Section \ref{weka}. Column 2 gives
the total number of significant internal parameters for that algorithm. Column 3 gives the processing time for each algorithm (running on a 64 bits PC AMD Phenom X3 8650 triple-core processor - 2.3GHz), 
and Columns 4, 5 and 6 give the quantities
$\langle\mbox{Compl}\rangle_{\mbox{\scriptsize{bright}}}$,
$\langle\mbox{Compl}\rangle_{\mbox{\scriptsize{faint}}}$ and
$\mbox{Compl}_{20.75}$ along with their standard deviations. The rows
in Table \ref{tab.parameter_space} have been ordered according to
the values of $\langle\mathrm{Compl}\rangle_{\mathrm{faint}}$.

It is clear from Table \ref{tab.parameter_space} (and Figure 2) that all of the algorithms have comparable efficiency at brighter magnitudes ($r < 19$). However, at $r \ge 19$ their performance varies significantly. The Decision Stump, with no internal parameters,  unsurprisingly performs worst, although it is the fastest algorithm. The NBTree, which also has no internal parameters, is considerably better albeit more computationally expensive. The fast algorithms, including Simple Cart, J48 and J48graft, REPTree and  Random Tree, have mean faint-end completeness of $\langle\mbox{Compl}\rangle_{\mbox{\scriptsize{faint}}} \sim 81\%-83\%$ but with $\mbox{Compl}_{20.75} < 70\%$. The remaining algorithms provide better faint end completeness at the cost of increasing processing times. Note that all of the algorithms are almost equally robust, as measured from the dispersions of their mean completeness. We conclude that the FT algorithm is optimal because of its greater accuracy at faint magnitudes while still requiring only modest processing time. In fact, as shown in Table \ref{tab.parameter_space}, the FT algorithm is not only the most accurate among the 13 we tested, but also very robust (ranked second in dispersion of mean completeness).

To examine what causes the different success rates among algorithms, we compare the best (FT) and worst (Random Tree) performing DTs. We examined the attribute values for objects where the classifications from the two methods agreed and where they disagreed. We find that there are some attributes where the objects for which the two classifiers agree and for which they disagree are clearly separated. In the FT vs. Random Tree comparison, the separation is greatest in Petrosian attributes (especially PetroR90 and petroRad). The exact reasons for this are unclear, but because each algorithm uses the same attributes, it must be an artifact of the tree construction. This further reinforces the need to extensively test classifiers with large training samples to find the best algorithm.

\begin{table*}
\caption{Main results of the comparative study of WEKA algorithms. The columns are the name of the algorithm tested; the number of parameters of the algorithm that can change the resultant tree; the mean processing time averaged over all parameter sets tested; the mean completeness averaged over the magnitude interval $[14,19]$; the mean completeness averaged over the magnitude interval $[19,21]$; and the completeness in the faintest magnitude bin ($20.5\le r \le 21.0$)}
\begin{center}
\begin{tabular}{lccccc}\hline
\hline \multicolumn{1}{c}{} Algorithm &  Number of  & Processing Time & $\langle\mbox{Compl}\rangle_{\mbox{\scriptsize{bright}}}$ & $\langle\mbox{Compl}\rangle_{\mbox{\scriptsize{faint}}}$ & $\mbox{Compl}_{20.75}$ \\
       \multicolumn{1}{c}{}             & Parameters  &     (hours)     & $\%$ & $\%$ & $\%$ \\
\hline  
\hline  
         Decision Stump & $0$ & $0.03$    & $99.20(\pm0.17)$ & $68.06(\pm1.20)$ & $42.29(\pm9.64)$ \\
        NBTree        & $0$ & $1.12$    & $99.64(\pm0.16)$ & $79.19(\pm1.39)$ & $63.55(\pm14.90)$ \\
        J48graft      & $4$ & $0.09$   & $99.74(\pm0.12)$ & $80.93(\pm1.16)$ & $65.84(\pm10.39)$ \\
        Simple Cart    & $3$ & $0.05$   & $99.63(\pm0.16)$ & $81.56(\pm1.13)$ & $67.06(\pm9.51)$ \\
	J48           & $6$ & $0.08$   & $99.73(\pm0.12)$ & $81.70(\pm0.96)$ & $66.30(\pm7.69)$ \\
        REPTree       & $4$ & $0.09$   & $99.50(\pm0.18)$ & $82.76(\pm1.09)$ & $69.32(\pm8.80)$ \\
        Random Tree    & $2$ & $0.06$   & $99.50(\pm0.18)$ & $82.76(\pm1.09)$ & $69.32(\pm8.80)$ \\
        Random Forest  & $3$ & $1.13$   & $99.77(\pm0.12)$ & $83.15(\pm1.14)$ & $70.48(\pm9.91)$ \\
        BFTree        & $5$ & $0.24$   & $99.69(\pm0.15)$ & $83.18(\pm1.10)$ & $69.85(\pm9.55)$ \\
        ADTree        & $2$ & $1.42$   & $99.73(\pm0.12)$ & $83.80(\pm1.12)$ & $71.88(\pm9.81)$ \\
        LMT           & $2$ & $5.50$   & $99.66(\pm0.15)$ & $83.91(\pm1.14)$ & $72.18(\pm9.39)$ \\
        LADTree       & $1$ & $7.90$ & $99.70(\pm0.14)$ & $84.39(\pm1.10)$ & $72.74(\pm9.41)$ \\
        FT            & $3$ & $2.50$   & $99.64(\pm0.15)$ & $84.98(\pm1.08)$ & $74.04(\pm8.45)$ \\  
\hline
\label{tab.parameter_space}
\end{tabular}
\end{center}
\end{table*}


\subsection{Constructing The Final Decision Tree}
\label{dr7sgseparation}

\begin{figure}
\begin{center}
\includegraphics[width=80mm,height=80mm]{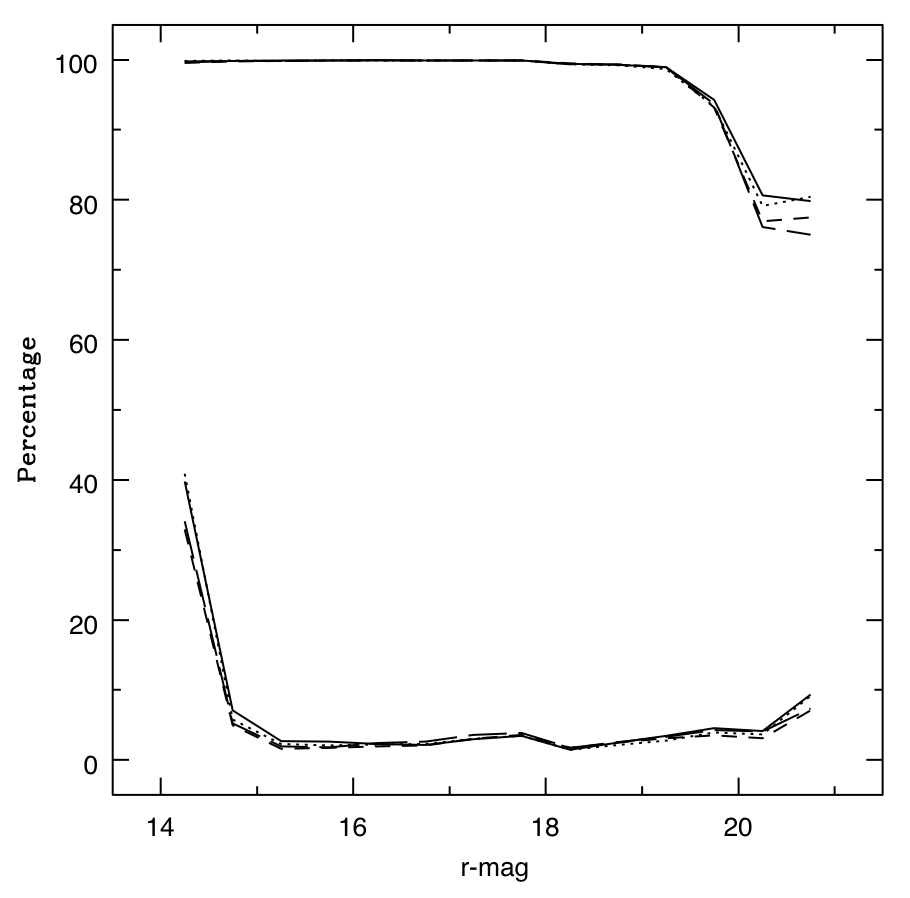}
\end{center}
\caption{The completeness (upper curves) and contamination (lower curves) functions for all four perturbed data sets used to train a DT with the FT algorithm. Each data set is represented by a different line type.}
\label{fig.perturb_fcompl}
\end{figure}

Having found that the FT algorithm provides the best star/galaxy separation performance in cross-validation tests, we must choose a training set to construct the final DT to classify objects from the SDSS-DR7 photometric catalog. As described in Section \ref{Data}, the CAS database provides 884,126 objects with star or galaxy spectral classification, 
which comprises our full training sample. However, using this entire data set within the WEKA implementation of the FT algorithm requires large amounts of computer memory, decreasing the overall performance. To see if the final tree depends strongly on the training set size and the exact values of each object's atrributes, we performed a test using subsets of the training data while perturbing the object attributes.
For each photometric attribute discussed in Section $4.1$, we generate a perturbed value,
\begin{equation}
 \label{eq.perturbation}
 X = X_{obs} + \sigma u ~\mbox{ ,}
\end{equation}
where $X_{obs}$ is the observed attribute value and $u$ is a random Gaussian deviate, 
with the dispersion $\sigma$ computed from the first and fourth quartiles of the attribute's uncertainty distribution. We then subdivided the training data with these perturbed attributes into four samples, each with $221,029$ objects, and built 4 different DT's. We required that each subset have the same class distribution as the full training set. We then use these four DT's to classify the original training set. The results are shown in Figure \ref{fig.perturb_fcompl}, which shows that the completeness function is unchanged in the magnitude range $14 \le r < 20$. At the faintest magnitudes, $20 \le r \le 21$ the completeness function varies slightly, by  $\lesssim5\%$. These results suggest that we can safely reduce the training set size by a factor of a few with no significant loss of accuracy. With this reduced training data set the DT construction is speeded up considerably. This test simultaneously confirms that the final DT is insensitive to measurement errors on the individual object attributes.

We further tested the dependence of DT classification success on training set size. In future optical surveys, we expect that spectroscopic data for the faintest objects will be at a premium, available for only a small fraction of the photometric objects. At $r>19$, the fainter part of our training sample, we constructed DTs using different sized subsamples of the available training data, ranging from 10\% to 100\% of the full training set. We find that the completeness remains essentially unchanged as long as at least 20\% of the faint-end training sample is used (about 7100 objects). This result suggests that future deep surveys may be able to perform accurate star/galaxy separation even with a modest truth sample for training, as long as there is sufficient resolution in the images.

To select data to train the final DT, we consider bright and faint objects separately. 
At $14 \le r < 19$ we selected 1/4 of the objects from the spectroscopic sample, maintaining the same magnitude distribution as in the full sample, yielding 205,348 objects. For magnitudes $19 \le r \le 21$ we kept all objects from the training set. These faint objects are poorly sampled spectroscopically, so we attempted to provide all possible information to the FT algorithm to maintain the completeness function at faint magnitudes. 

The final training set contains 240,712 objects with 13 attributes each, extracted from the original spectroscopic sample. The resultant decision tree was then applied to classify objects with $r$ magnitudes between 14 and 21 from the entire SDSS-DR7 Legacy survey catalog. A portion of the resulting catalog is shown in Table ~\ref{tabclass}; the full catalog is available as an electronic table. Column 1 gives the unique SDSS ObjID and Column 2 contains the SDSS \texttt{ModelMag} magnitude in $r$-band. Columns 3 and 4 give \texttt{$type_r$} and \texttt{type}, the SDSS classifications using $r$-band alone and using all five bands, respectively. The SDSS classifications are 0=unknown, 3=Galaxy, 6=Star, 8=Sky. Column 5 provides our FT Decision Tree classification, where 1 is a star and 2 is a galaxy. We note that we classified {\em all} objects with $14<$\texttt{$ModelMag_r$}$<21$, regardless of their SDSS classification. Detections which the SDSS has classified as \texttt{type}=0 or 8 should be viewed with caution as there is a high likelihood that they are not truly astrophysical objects.

\begin{table*}
\caption{Star/galaxy classification provided by SDSS and by our FT Decision Tree.  Column 1 lists the unique SDSS ObjID, Column 2 contains the SDSS modelMag magnitude in $r$-band. Columns 3 and 4 give $type_r$ and $type$, the SDSS classifications using $r$-band alone and using all five bands, respectively. Column 5 provides our FT Decision Tree classification, where 1 is a star and 2 is a galaxy. }
\begin{center}
\begin{tabular}{lcccc}\hline
SDSS ObjID &  ModelMag$_r$  & Type$_r$ & Type & FT Class \\
\hline  
588848900971299281 & 20.230947 & 3 & 3 & 2 \\
588848900971299284 & 20.988880 & 3 & 3 & 2 \\
588848900971299293 & 20.560146 & 3 & 3 & 2 \\
588848900971299297 & 19.934738 & 3 & 3 & 2 \\
588848900971299302 & 20.039648 & 3 & 3 & 2 \\
588848900971299310 & 20.714321 & 3 & 3 & 2 \\
588848900971299313 & 20.742567 & 3 & 3 & 2 \\
588848900971299314 & 20.342773 & 3 & 3 & 2 \\
588848900971299315 & 20.425304 & 3 & 3 & 2 \\
588848900971299331 & 20.582634 & 3 & 3 & 2 \\
\hline
\label{tabclass}
\end{tabular}
\end{center}
\end{table*}

In the next section, we present a comparative study of our catalog and other catalogs providing star/galaxy classification for the SDSS.

\section{Comparison With Other SDSS Classifications}
\label{comparativestudy}

We compare the results of our DT classification of objects in the SDSS photometric catalog with those from the parametric classifier used in the SDSS pipeline, the 2DPHOT software for image processing (La Barbera et al, 2008) and the DT classification from Ball et al. (2006). These comparisons utilize only objects with spectroscopic classifications so that the true classes are known. All three methods give classifications other than just stars or galaxies. However, as we are interested only in stars and galaxies, all samples described in this section are exclusively composed of objects which were classified by the respective methods as star or galaxy. 

\subsection{FT Algorithm Versus 2DPHOT Method}
\label{2dphot}
\begin{figure}
\centering
\includegraphics[width=80mm,height=80mm]{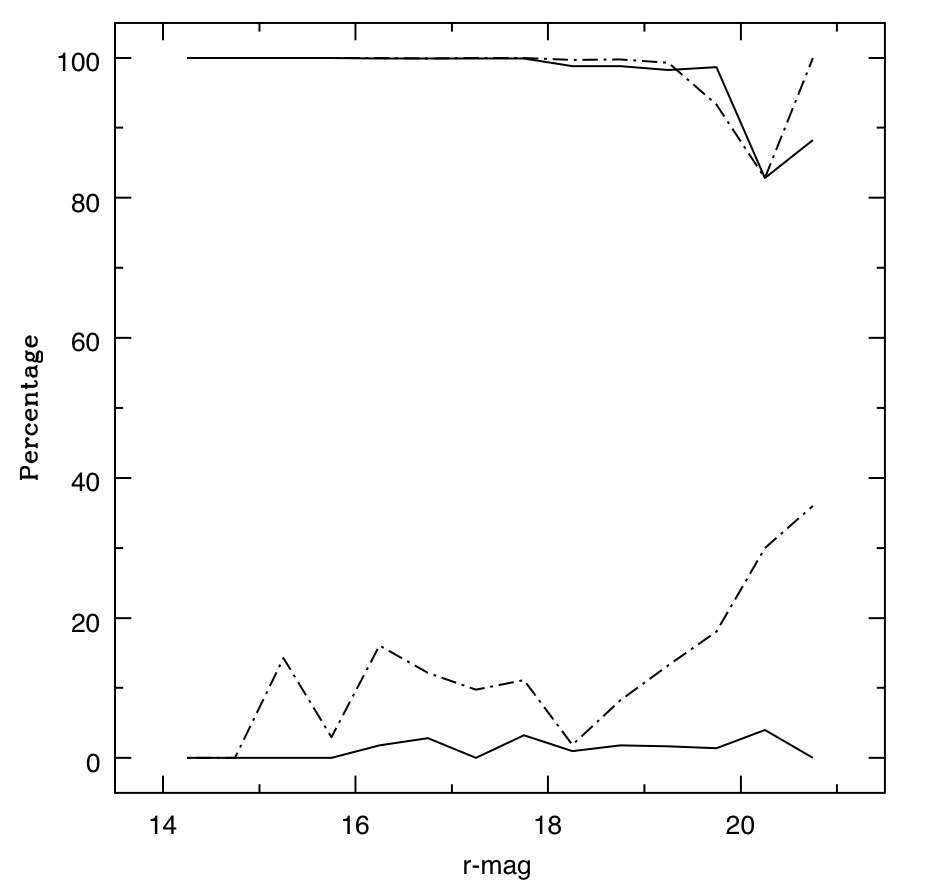}
\caption{Completeness (upper curves) and contamination (lower curves) for the sample of 10,391 SDSS objects reprocessed with 2DPHOT and classified as either star or galaxy.
The full lines show the completeness and contamination functions from our DT classification whereas the dash-dot lines are for the 2DPHOT classification.}
\label{fig.spec_2dphot_completeness}
\end{figure}

2DPHOT is a general purpose software package for automated source detection and analysis in deep wide-field images. It provides both integrated and surface photometry of galaxies in an image and performs star-galaxy separation by defining a stellar locus in its parametric space (La Barbera et al. 2008). The comparison was done for $10,391$ objects from the spectroscopic sample which have been reprocessed by 2DPHOT, with the results presented in Figure \ref{fig.spec_2dphot_completeness}. 
We see that both classifiers have the same completeness trends with magnitude, but our FT classifier generates almost no contamination, while contamination in 2DPHOT reaches $\sim 40\%$. 

\subsection{FT Algorithm Versus Axis-Parallel DT}
\label{ball}
\begin{figure}
\centering
\includegraphics[width=80mm,height=80mm]{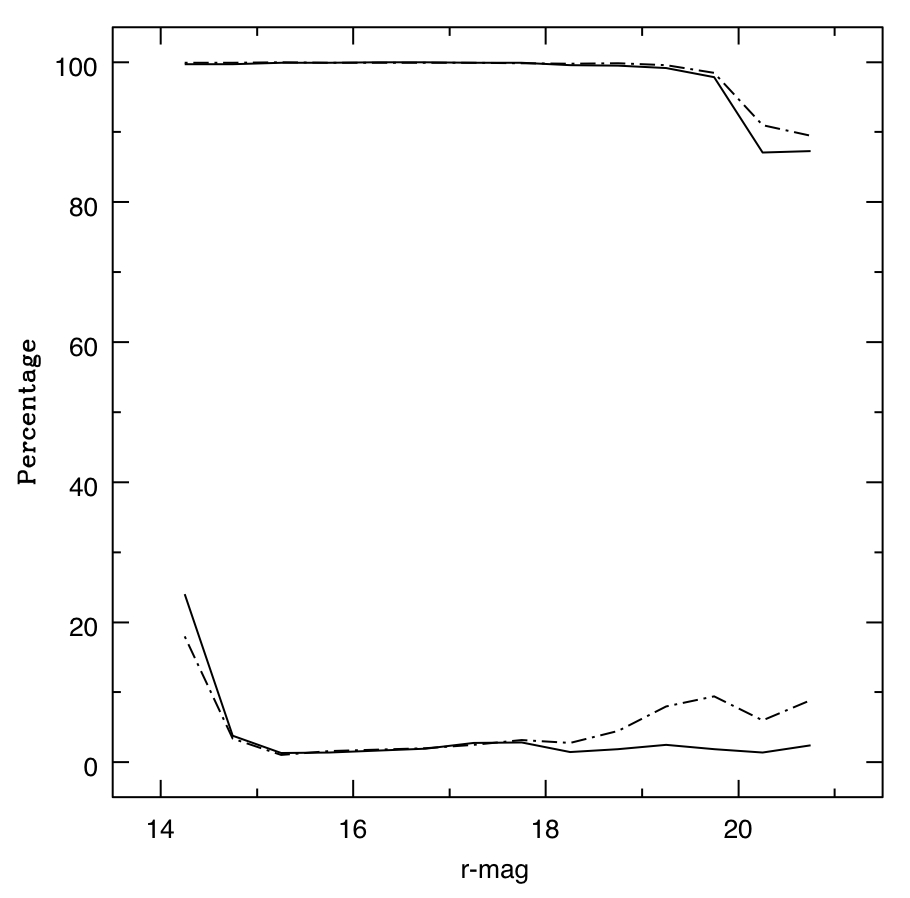}
\caption{Completeness (upper curves) and contamination (lower curves) for the $561,070$ objects with SDSS spectroscopic data processed by Ball et al. (2006). The full lines give the completeness and contamination functions of our DT classification whereas the dash-dot lines give the same for Ball's classification.}
\label{fig.ball_our_spec}
\end{figure}

Ball et al. (2006) were the first to apply the DT methodology to SDSS data. They use an axis-parallel DT to assign a probability that an object belongs to one of three classes: stars, galaxies or nsng (neither star nor galaxy). The completeness and contamination functions for both Ball's axis-parallel DT and for our FT, calculated using a sample of $561,070$ objects from Ball's catalog, are shown in Figure \ref{fig.ball_our_spec}. 
These results show that our FT tree perform similarly to the axis-parallel tree, but the FT generates lower contamination than the axis-parallel tree. At the faint end ($r\ge19$) our contamination remains constant around $3\%$ while Ball's catalog has a contamination rate of $\sim9\%$. 

\subsection{FT Algorithm Versus SDSS Pipeline Parametric Method}
\label{photo}
The SDSS pipeline classifies an object into 3 classes: unknown, galaxy or star using the difference between the psf and model magnitudes (cf. Section \ref{Attributes}). If the condition $psfMag - modelMag > 0.145$ is satisfied, the object is classified as galaxy; otherwise, it is classified as a star. 

\begin{figure}
\begin{center}
\includegraphics[width=80mm,height=80mm]{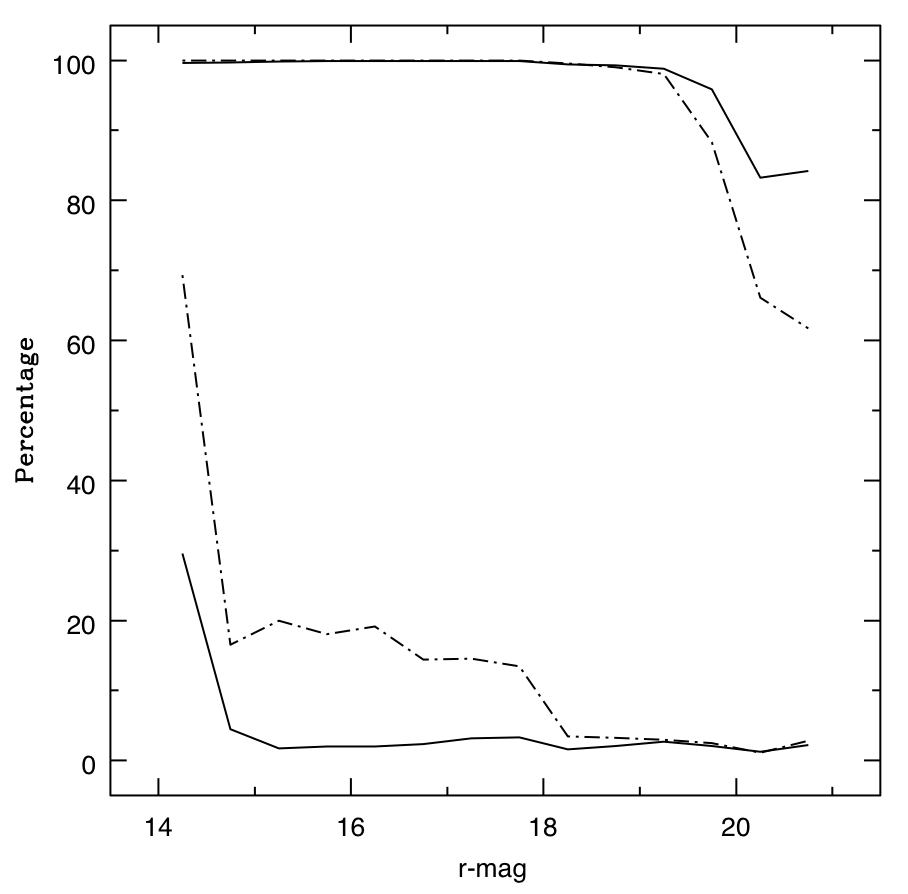}
\end{center}
\caption{The completeness (upper curves) and contamination (lower curves) functions for the SDSS parametric method (dash-dotted lines) and for our DT method (solid lines), applied to the 880,715 objects of the spectroscopic sample (see text).}
\label{fig.sdss_our_spec}
\end{figure}

We first analyzed the behavior of our final DT in comparison with that of the SDSS pipeline for the full spectroscopic sample. We note that this will contain the same objects used to initially train the DT, as discussed in Section \ref{dr7sgseparation}. However, because  only $\sim$ 25\% of the entire sample is used for training, we neglect its influence on the results. The completeness and contamination curves are shown in Figure \ref{fig.sdss_our_spec} for a sub-sample of the entire spectroscopic sample where $8,406$ objects with at least one saturated pixel have been removed. The results show that the completeness of our DT classification stays above $80\%$ for magnitudes $r \ge 19$ with negligible stellar contamination. In contrast, the completeness for the SDSS pipeline drops to $60\%$ in the same range, and is highly contaminated at the brighter magnitudes. These results show that application of our DT provides a gain in completeness of $\sim 20\%$ at faint magnitudes when compared with the SDSS pipeline. Our FT tree also yields much lower contamination than the SDSS parametric method; while the FT contamination stays around $6\%$, the SDSS parametric contamination is about $17\%$ for magnitudes brighter than $18$.

\begin{figure}
\begin{center}
\includegraphics[width=80mm,height=80mm]{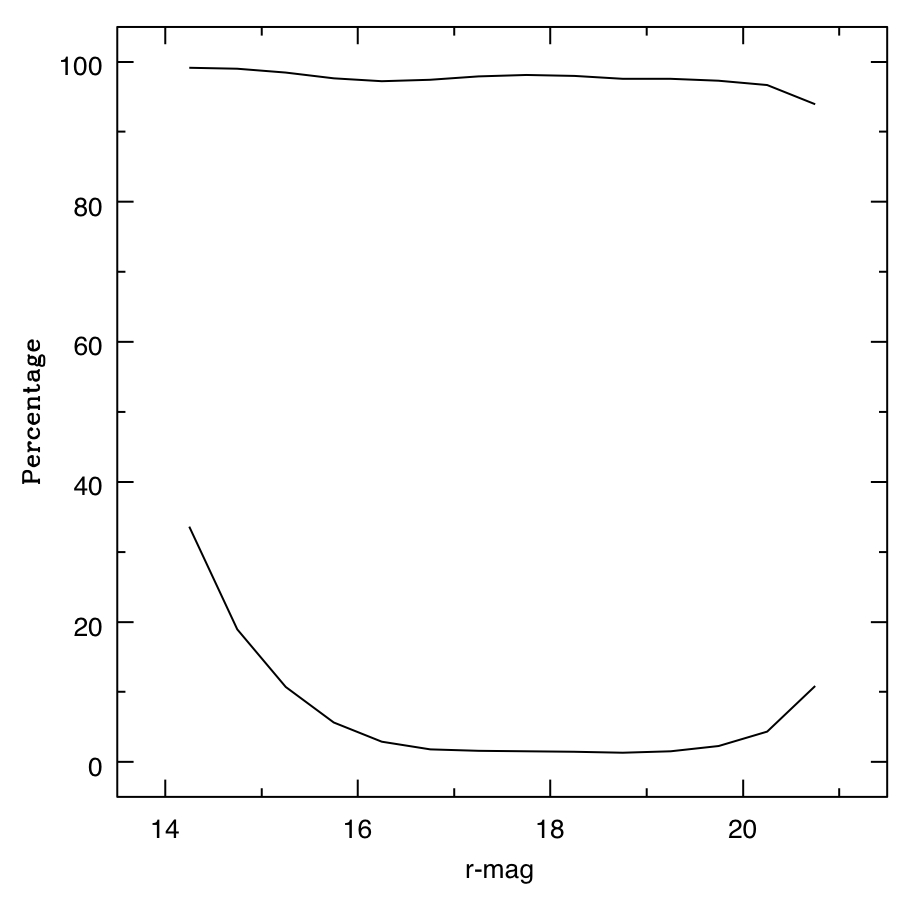}
\end{center}
\caption{The completeness and contamination for the SDSS parametric method when our DT classification is taken as the truth.}
\label{fig.sdssphot_FT}
\end{figure}

Finally we compared our DT classifications and the SDSS parametric classifications for all objects in the application sample (see Section \ref{Data}). Since there is no {\it a priori} true classification in this case (unlike the training sample), we compare these 2 methods by assuming that the DT is correct, based on its better performance (described above). In Figure \ref{fig.sdssphot_FT}  we show the completeness and contamination functions for the SDSS parametric classification assuming that our DT classification is correct. Only objects that are classified as either star or galaxy by both the SDSS parametric method and the FT DT are considered here, with other classifications in SDSS being an irrelevant fraction ($0.05\%$ of the total sample). We see from this figure that there is significant disagreement between our DT classification and the SDSS parametric method when classifying stars, implying a greater stellar contamination in the SDSS classification. The completeness shows, at faint magnitudes $20.5\le r\le21$, that the two classifiers disagree on $\sim 6\%$ of the whole application sample.

\subsection{A Simple Test of the SDSS Pipeline Parametric Method}
\label{photo}

The SDSS parametric method relies on a single parameter test: if the condition $psfMag - modelMag > 0.145$ is satisfied, the object is classified as galaxy; otherwise, it is classified as a star. The cutoff between the two classes was chosen based on simulated images of stars and galaxies. However, the choice of dividing line between stars and galaxies in this parameter space can be improved using the extenssive spectroscopic training sample. To test this, we retrieved $psfMag - modelMag$ for the training sample, and used the Decision Stump to generate a single node Decision Tree. In creating such a tree, the Decision Stump will seek the value of $psfMag - modelMag$ which maximizes completeness while minimizing contamination. Surprisingly, we find that the optimal requirement for an object to be classified as a galaxy is $psfMag - modelMag > 0.376$, significantly different from the value used by SDSS.

To see why there is such a large difference, we have examined images of bright objects misclassified by the SDSS parametric method, which are responsible for the high contamination rate shown in Figure 6. We find that many of these objects are in close pairs. Example of nine such objects are shown in Figure 8, where each panel includes an object misclassified by the SDSS parametric method. We clearly see that many are either heavily blended with a companion or close enough to another object for the value of $psfMag-modelMag$ used by the SDSS parametric classifier to be influenced by the neighboring object. This is clearly visible in Figure 9, which shows $psfMag - modelMag$ as a function of magnitude for the training sample. Spectroscopically classified stars are shown in red, while galaxies are shown in green. A second ridgeline of stars is clearly seen with $psfMag - modelMag \sim 0.3$, all of which are misclassified with the standard SDSS parametric cut. This issue was noted already in the SDSS Early Data Release (Stoughton et al. 2002), where they noted that un-deblended star pairs and galaxies with bright nuclei are improperly classified. This further validates our optimal DT classifier, which, by using a larger number of attributes and not combining them, is better able to create a rule set for correctly classifying even those objects with nearby companions.

\begin{figure}
\centering
\includegraphics[width=70mm,height=70mm]{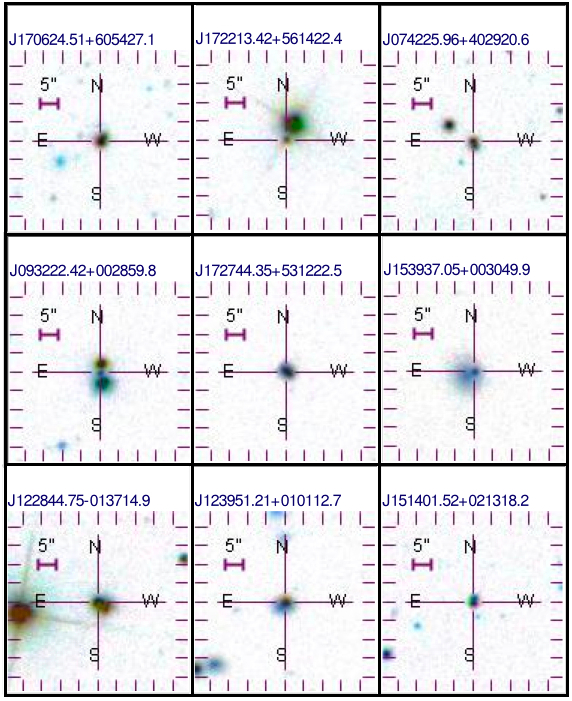}
\caption{Postage stamps from SDSS DR7 for nine objects which are misclassified by the SDSS parametric method. Almost all are blends or have brighter nearby companions which affect the photometry.}
\label{blends}
\end{figure}

\begin{figure}
\centering
\includegraphics[width=80mm,height=80mm]{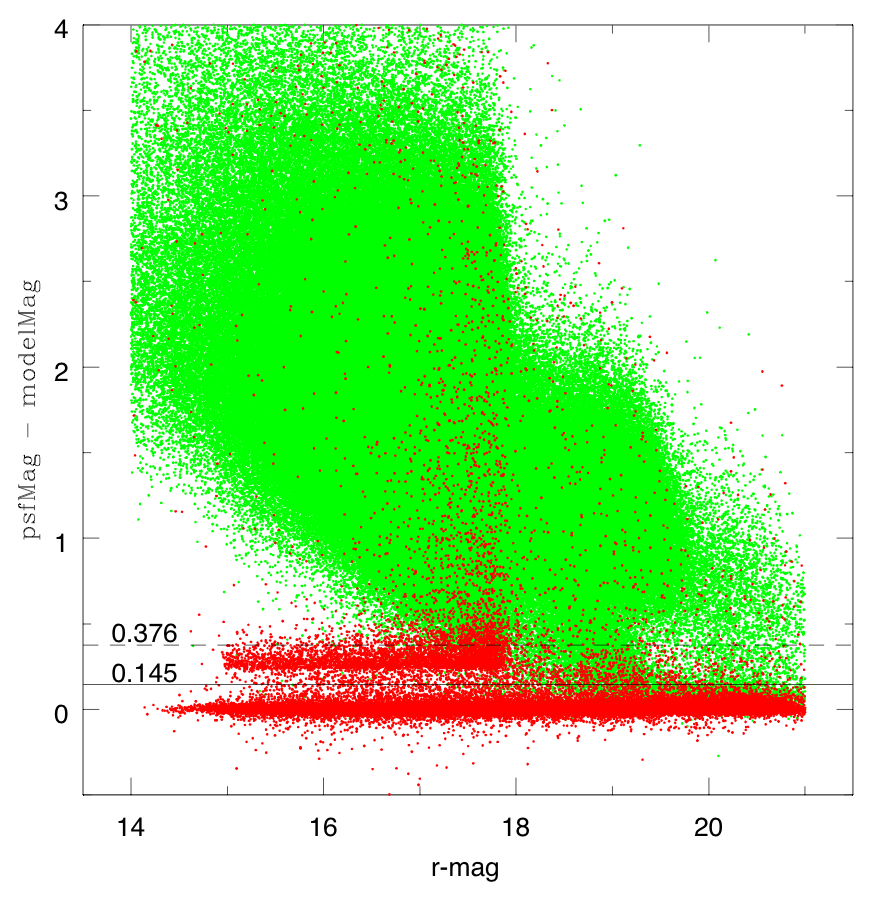}
\caption{The $psfMag - modelMag$ parameter used in the standard SDSS classifier is plotted as a function of magnitude for the spectroscopic training sample. Stars are shown as red dots, while galaxies are in green. The dividing lines used by the SDSS classifier ($psfMag - modelMag=0.145$) and derived by the Decision Stump ($psfMag - modelMag=0.376$) are also shown. The SDSS classifier incorrectly assigns galaxy classifications to many relatively bright stars, most of which have nearby neighbors.}
\label{sdssdecstump}
\end{figure}

We note that it is possible that there exist attributes in the SDSS database that we do not employ but which could improve classification accuracy when used in a DT. It may also be the case that PSF-deconvolved attributes, such as those measured by 2DPHOT, can improve performance. Other ``indirect'' attributes, such as colors or combinations of multiple attributes, can also be considered, as done by Ball et al. and the SDSS parametric method. However, testing all possible sets of object attributes is outside the scope of this work. Other avenues for further testing include generating a Committee Machine, which looks at the outputs of multiple classification algorithms and chooses a final meta-class for each object based on majority vote or some other criterion. Nevertheless, our DT significantly outperforms all other published classifiers applied to SDSS photometry.

\section{Summary}
\label{summary}

We analyzed the star/galaxy separation performance of 13 different decision tree algorithms from the publicly available data mining tool WEKA when applied to SDSS photometric data. This is the first examination of a public data mining tool applied to such a large catalog. 
We show the completeness and contamination functions for all of the algorithms in an extensive study of each algorithm's parameter space. These functions were obtained using cross-validation tests and demonstrate the capability of each algorithm to classify objects from training data. Thus, our study may be used as a guide for astronomers that desire to apply such data mining algorithms in star-galaxy separation tasks and other similar data mining problems. The main results of our work are:
\begin{enumerate}
 \item { 13 different algorithms from WEKA are tested and Figure 2 shows the locus of the resultant completeness functions;}
 \item {All algorithms achieve the same accuracy in the magnitude range $14\le r <19$, but with large differences in the required processing time (Table \ref{tab.parameter_space} and Figure 2);}
 \item {The completeness functions in the faint magnitude interval ($r\ge 19$) show that the ADTree, LMT and FT are the most robust and have similar performance (see Table \ref{tab.parameter_space}). However, FT requires approximately half the time to build a DT than the others.}
 \item {The WEKA FT algorithm is therefore chosen as the optimal DT for classifying SDSS-DR7 objects based on photometric attributes;}
 \item {We show, using this FT, that reducing the size of the training data by a factor of $\sim5$ does not change significantly the completeness and contamination functions (see Figure \ref{fig.perturb_fcompl});}
 \item {We use the FT WEKA algorithm to construct a DT trained with photometric attributes and spectral classifications for $240,712$ objects from the SDSS-DR7, and apply this DT to separate stars from galaxies in the Legacy survey sample of objects in the magnitude range $14\le r \le21$ from SDSS-DR7;}
 \item {Finally, we compare our results with the SDSS parametric method, 2DPHOT and Ball's axis-parallel DT. Our catalog has much lower contamination than all three methods (Figures \ref{fig.spec_2dphot_completeness}, \ref{fig.ball_our_spec} and \ref{fig.sdss_our_spec}), and a higher completeness than the SDSS parametric method for faint objects ($r\ge19$).}
\end{enumerate}

\begin{acknowledgements}
We thank Robert Hanisch for helpful comments and suggestions and
Nick Ball for providing data for comparison with his star/galaxy separation based on SDSS-DR3.
Funding for the  SDSS  and SDSS-II  has  been provided  by  the  Alfred P.   Sloan
Foundation,  the  Participating  Institutions,  the  National  Science
Foundation, the  U.S.  Department of Energy,  the National Aeronautics
and Space Administration, the  Japanese Monbukagakusho, the Max Planck
Society, and  the Higher Education  Funding Council for  England.  The
SDSS Web  Site is  http://www.sdss.org/.  The SDSS  is managed  by the
Astrophysical Research Consortium  for the Participating Institutions.
The  Participating Institutions  are  the American  Museum of  Natural
History,  Astrophysical   Institute  Potsdam,  University   of  Basel,
University of  Cambridge, Case Western  Reserve University, University
of Chicago,  Drexel University,  Fermilab, the Institute  for Advanced
Study, the  Japan Participation  Group, Johns Hopkins  University, the
Joint  Institute for  Nuclear  Astrophysics, the  Kavli Institute  for
Particle Astrophysics  and Cosmology, the Korean  Scientist Group, the
Chinese Academy of Sciences  (LAMOST), Los Alamos National Laboratory,
the     Max-Planck-Institute     for     Astronomy     (MPIA),     the
Max-Planck-Institute   for  Astrophysics   (MPA),  New   Mexico  State
University,   Ohio  State   University,   University  of   Pittsburgh,
University  of  Portsmouth, Princeton  University,  the United  States
Naval Observatory, and the University of Washington.
\end{acknowledgements}


\begin{thebibliography}{99}
\bibitem[Abazajian et al.(2004)]{abazajian2004} Abazajian, K., et al. 2004, AJ, 128, 502
\bibitem[Abazajian et al.(2005)]{abazajian2005} Abazajian, K., et al. 2005, AJ, 129, 1755
\bibitem[Abazajian et al.(2009)]{abazajian2009} Abazajian, K., et al. 2009, ApJ Supplement Series, 182, 543
\bibitem[Ball et al.(2006)]{ball2006} Ball, N. M., et al. 2006, ApJ, 650, 497
\bibitem[Bernstein \& Jarvis (2002)]{bernstein2002}Bernstein, G. M., Jarvis, M. 2002, AJ, 123, 583
\bibitem[Breiman (2001)]{breiman2001}Breiman, L. 2001, Machine Learnin, 45, 5
\bibitem[Breiman et al. (1984)]{breiman1984}Breiman, L., et al. 1984, Classifications and Regression Trees, Wadsworth Internetworkional Group
\bibitem[(Fayyad, 1994)]{fayyad1994}Fayyad, U. M. 1994, in Proc. of the Twelfth National Conference on Artificial Intelligence, 1, 601
\bibitem[(Fayyad \& Irani, 1992)]{fayyad_irani1992}Fayyad, U. M., Irani, K. B. 1992, in proc. of the Tenth National Conference on Artificial Intelligence, 1, 104
\bibitem[Freund and Mason (1999)]{freund1999}Freund, Y. Mason, L. 1999, In Proceedings of the 16$^{th}$ International Conference on Machine Learning, 124. 
\bibitem[Geoffrey, 1999]{geoffrey1999}Geoffrey, I. W., Geelong, V. 1999, The $16^{\mbox{\tiny{th}}}$ International Joint Conferences on Artificial Intelligence, 2, 702
\bibitem[Haijian Shi (2007)]{haijian2007}Haijian Shi 2007, Master Thesis, Hamilton, NZ
\bibitem[Heydon-Dumbleton 1989]{heydon1989}Heydon-Dumbleton, N. H., et al.. 1989, MNRAS, 238, 379
\bibitem[Holmes et al. (2001)]{holmes2001}Holmes, G., et al. 2001, ECML, 161
\bibitem[Jo\~ao Gama (2004)]{joao2004} João Gama 2004, Machine Learning, 55(3), 219
\bibitem[La Barbera et al. (2008)]{2dphot}F. La Barbera, et al. 2008, PASP, 702, 120
\bibitem[MacGillivray et al.(1976)]{macgillivray1976}MacGillivray, H. T., et al. 1976, MNRAS, 176, 265
\bibitem[Maddox et al.()]{maddox1990}Maddox, S. J., et al. 1990, MNRAS, 243, 692
\bibitem[Odewahn et al.1999]{odewahn1999}Odewahn, S. C., et al. 1999,in BAAS, 31, 1235
\bibitem[Odewahn et al.2004]{odewahn2004}Odewahn, S. C., et al. 2004, AJ, 128, 3092
\bibitem[Petrosian (1976)]{petrosian1976}Petrosian, V. 1976, ApJ, 209, L1
\bibitem[Quinlam(1993)]{quinlam1993}Quinlam, J. R. 1993, C4.5: Programs for machine learning, Morgan Kaufmann
\bibitem[Quinlam(1986)]{quinlam1986}Quinlam, J. R. 1986, Machine Learning, 1, 81
\bibitem[Ruiz et al. (2009)]{ruiz2009} Ruiz, R. S. R., et al. 2009, TEMA, 10, 75
\bibitem[stoughton et al. (2002)]{stoughton2002} Stoughton, C., et al. 2002, AJ, 123, 485
\bibitem[Suchkov et al.(2005)]{suchkov2005}Suchkov , A. A., et al. 2005, AJ, 130, 2401
\bibitem[Weir et al.1995]{weir1995}Weir, N. et al. 1995, ApJ, 109, 2401
\bibitem[Witten \& Franck (2000)]{witten2000} Witten, I. H. \& Frank, E. 2000, Data Mining: practical machine learning toos and techniques with java implementations(San Fancisco:Morgan Kaufmann)
\bibitem[Yasuda et al.(2001)]{YASU2001} Yasuda, N. et al. 2001, AJ, 122, 1104
\bibitem[York et al.(2000)]{york2000}York, D. G., et al. 2000, AJ, 120, 1579
\end{thebibliography}
\end{document}